\documentclass[aps,twocolumn,reprint,preprintnumbers,showpacs,10pt]{revtex4-2}

\usepackage[english]{babel}

\usepackage{amsmath}
\usepackage{comment}
\usepackage{graphicx}
\usepackage[colorlinks=true, allcolors=blue]{hyperref}

\usepackage{tikz,subfigure}
\usepackage{dcolumn}
\usepackage{amsmath,booktabs}
\usepackage{amsthm,amssymb}
\usepackage{amstext}
\usepackage{orcidlink}
\usepackage{mathrsfs,booktabs} 
\usepackage{graphicx,subfigure,setspace,makecell}
\usepackage[normalem]{ulem}
\newcommand{\dd}{\mathrm{d}} 
\newcommand{\ee}{\mathrm{e}}

\usepackage{oplotsymbl} 

\newcommand{\ZW}[1]{ \textcolor{orange}{ZW: #1}}

\allowdisplaybreaks

\begin{document}
\preprint{MPP-2025-232, USTC-ICTS/PCFT-25-61, LAPTH-056/25}
\title{Complete computation of all three-loop five-point massless planar integrals}
\author{Dmitry Chicherin$^{a}$}
\email{chicherin@lapth.cnrs.fr}
\author{Yu Wu$^{b}$}
\email{wy626@mail.ustc.edu.cn}
\author{Zihao Wu\,\orcidlink{0000-0003-3561-5403}$^{c,d,e}$}
\email{wuzihao@mail.ustc.edu.cn}
\author{Yongqun Xu\,\orcidlink{0000-0001-7775-3498}$^{b}$}
\email{yongqunxu@mail.ustc.edu.cn}
\author{Shun-Qing Zhang\,\orcidlink{0000-0002-0129-8959}$^{f}$}
\email{sqzhang@mpp.mpg.de}
\author{Yang Zhang$^{b,g,h}$}
\email{Corresponding author: yzhphy@ustc.edu.cn}

\affiliation{
$^a$ LAPTh, Universit\'e Savoie Mont Blanc, CNRS, B.P. 110, F-74941 Annecy-le-Vieux, France\\
$^b$ Interdisciplinary Center for Theoretical Study, University of Science and Technology of China, Hefei, Anhui 230026, China\\
$^c$ State Key Laboratory of Nuclear Physics and Technology, Institute of Quantum Matter, South China Normal University, Guangzhou 510006, China\\
$^d$ Guangdong Basic Research Center of Excellence for Structure and Fundamental Interactions of Matter, Guangdong Provincial Key Laboratory of Nuclear Science, Guangzhou 510006, China\\
$^e$ Hangzhou Institute for Advanced Study,
University of Chinese Academy of Sciences (HIAS, UCAS)\\
$^f$ Max-Planck-Institut f\"{u}r Physik, Werner-Heisenberg-Institut, Boltzmannstr. 8,
85748 Garching, Germany \\
$^g$ Peng Huanwu Center for Fundamental Theory, Hefei, Anhui 230026, China\\
$^h$ Center for High Energy Physics, Peking University, Beijing 100871, People’s Republic of China
}

\begin{abstract}
We  calculate all three-loop, five-point, massless planar Feynman integral families in the dimensional regularization scheme. This is a new milestone in Feynman integral computations. The analysis covers four distinct families of Feynman integrals for this configuration, for all of which we derive the canonical differential equations. Our results also confirm a prediction on the three-loop five-point alphabet. The boundary values are analytically determined. Using these differential equations, the integrals can be evaluated to high precision efficiently. Our work establishes the foundation for next-to-next-to-next-to-leading-order (N$^3$LO) calculation of the production of three massless final states, as well as corresponding bootstrap studies in gauge theories.  
\end{abstract}

\maketitle

\section{Overview}
To study the nature of our world at both microscopic and macroscopic levels is the main goal of modern physics. In the microscopic sense, large colliders serve as ``microscopes'' to study the properties of the fundamental particles and their interactions over the past decades. In the colliders, huge amounts of data are produced as the detectors probe the production of the collision. As these data accumulate, there are higher demands for theoretical computations in quantum field theory (QFT), in order to predict the collision events at a high precision level. In the macroscopic sense, gravitational waves (GW) are crucial for probing the universe. In order to observe GW by detectors, precise templates are needed. High-precision computation also provides an analytical approach to the computation of these templates at high-precision level. 
In this article, we report a new breakthrough on the direction of high-precision computation in QFT: the analytical results of all planar three-loop five-point Feynman integral families. 
These results open the door of the theoretical predictions for collision processes with three end states at the precision of the next-to-next-to-next-to-leading-order (N$^3$LO).
We also expect that the state-of-the-art techniques used in this work can make the study of GW physics from QFT approach more feasible.

\section{Introduction}

Precise computation in quantum field theory becomes an important task in recent decades. 
Evaluating the scattering amplitudes and their associated multi-loop Feynman integrals is crucial to collider phenomenology, theoretical studies of gauge theories. It also recently enhances the GW precision physics. On the experimental side, the development of future colliders-- the High-Luminosity Large Hadron Collider (HL-LHC)~\cite{ZurbanoFernandez:2020cco}, Future Circular Collider (FCC)~\cite{FCC:2024lyi}, Circular Electron Positron Collider (CEPC)~\cite{CEPCStudyGroup:2023quu}, International Linear Collider (ILC)~\cite{ILC:2007bjz} and future gravitational wave observatories--Laser Interferometer Space Antenna (LISA) \cite{LISA:2024hlh}, the Taiji \cite{Ruan:2018tsw}
and Tianqin \cite{TianQin:2015yph} projects, place unprecedented demands on theoretical accuracy. On the theoretical side, recent years have witnessed substantial progress in the precise theoretical computations, represented by two-loop Feynman integrals and the relevant next-to-next-to leading order (NNLO) amplitudes for phenomenology, most notably in the computation of two-loop five-point to several external mass~\cite{Gehrmann:2015bfy,Gehrmann:2018yef,Abreu:2018aqd,Chicherin:2018old,Chicherin:2020oor,Canko:2020ylt,Chicherin:2021dyp,Kardos:2022tpo,Abreu:2023rco,Badger:2021nhg,Abreu:2021oya,Chawdhry:2021mkw,Agarwal:2021vdh,Czakon:2021mjy,Becchetti:2025qlu,Abreu:2023bdp,Badger:2024sqv,Abreu:2024yit,Badger:2024lft,Badger:2024fgb,Wang:2024pmv,Becchetti:2025qlu,Chicherin:2024hes,Badger:2024dxo,Badger:2024mir,Brancaccio:2024map,Badger:2025uym,Badger:2025ljy,Wang:2025kpk}, massless six-point integrals~\cite{Henn:2024ngj,Henn:2025xrc,Carrolo:2025pue,Carrolo:2025agz} and the associated scattering amplitudes and associated observables.

The above achievements are enabled by the improvements of techniques for loop integral computation~\cite{Smirnov:2019qkx,Klappert:2020nbg,Wu:2023upw,Wu:2025aeg,Lange:2025fba,Smirnov:2025prc}, including the reduction of the integrals, the construction and solution of canonical differential equations (CDE)~\cite{Henn:2013pwa,Henn:2014qga}, and corresponding mathematical techniques~\cite{Klappert:2019emp,Klappert:2020aqs}.

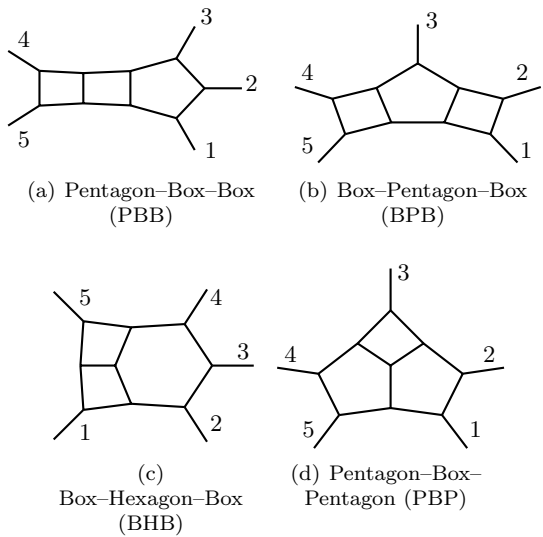
\begin{figure}[htbp]
		\centering
		\subfigure[Pentagon--Box--Box (PBB)
		] 
		{\begin{tikzpicture}[thick, scale=0.54]
				\coordinate (1) at (4.59314,0);
				\coordinate (2) at (5.75464,1.52085);
				\coordinate (3) at (4.59678,3.04424);
				\coordinate (4) at (0.00141206,2.44223);
				\coordinate (5) at (0,0.602005);
				\coordinate (6) at (4.14059,0.785867);
				\coordinate (7) at (4.83428,1.52169);
				\coordinate (8) at (4.14225,2.25834);
				\coordinate (9) at (0.771176,1.95129);
				\coordinate (10) at (0.770847,1.0941);
				\coordinate (11) at (3.00879,1.09988);
				\coordinate (12) at (1.85009,1.89538);
				\coordinate (13) at (3.00904,1.94518);
				\coordinate (14) at (1.85063,1.15019);
				\draw (8)--(13)(9)--(12)(11)--(13)(12)--(14)(6)--(11)(9)--(10)(11)--(14)(7)--(8)(12)--(13)(10)--(14)(6)--(7);
				\draw (1)--(6) (2)--(7)(3)--(8)(4)--(9)(5)--(10);
				\node[above] at  (5.99,1.22) {\small $2$};
				\node[above] at  (4.8871,2.85158) {\small $3$};
				\node[above] at  (0.386282,2.3) {\small $4$};
				\node[right] at  (4.6,0) {\small $1$};
				\node[below] at  (0.385335,0.7) {\small $5$};
			\end{tikzpicture}	\label{fig:PBB}}
		\subfigure[Box--Pentagon--Box (BPB)] {
			\begin{tikzpicture}[thick, scale=0.57]
				\coordinate (1) at (0.542344,0);
				\coordinate (2) at (0,1.76208);
				\coordinate (3) at (2.86764,3.22052);
				\coordinate (4) at (5.73463,1.76188);
				\coordinate (5) at (5.19173,0.000440965);
				\coordinate (6) at (1.17398,0.661902);
				\coordinate (7) at (0.874812,1.4885);
				\coordinate (8) at (2.86652,2.31261);
				\coordinate (9) at (4.85855,1.48833);
				\coordinate (10) at (4.55936,0.661564);
				\coordinate (11) at (2.2473,0.935929);
				\coordinate (12) at (3.48697,0.935849);
				\coordinate (13) at (1.90547,1.7419);
				\coordinate (14) at (3.82728,1.74175);
				\draw (8)--(13)(7)--(13)(9)--(14)(11)--(13)(12)--(14)(6)--(11)(9)--(10)(11)--(12)(8)--(14)(6)--(7)(10)--(12);
				\draw (1)--(6) (2)--(7)(3)--(8)(4)--(9)(5)--(10);
				\node[above] at  (5.3,1.7) {\small $2$};
				\node[above] at  (3.2,2.9) {\small $3$};
				\node[above] at  (0.3,1.7) {\small $4$};
				\node[below] at  (5.4,0.7) {\small $1$};
				\node[below] at  (0.3,0.8) {\small $5$};
			\end{tikzpicture}\label{fig:BPB}}
		\subfigure[Box--Hexagon--Box (BHB)] {
			\begin{tikzpicture}[thick, scale=0.60]
				\coordinate (1) at (0,3.3263);
				\coordinate (2) at (3.39874,3.37806);
				\coordinate (3) at (4.43267,1.68833);
				\coordinate (4) at (3.39699,0);
				\coordinate (5) at (0.000346146,0.0515417);
				\coordinate (6) at (0.658947,2.67226);
				\coordinate (7) at (2.90173,2.60935);
				\coordinate (8) at (3.51164,1.68889);
				\coordinate (9) at (2.90067,0.7698);
				\coordinate (10) at (0.658598,0.706904);
				\coordinate (11) at (1.36335,1.68989);
				\coordinate (12) at (0.592501,1.69016);
				\coordinate (13) at (1.71772,2.53747);
				\coordinate (14) at (1.71696,0.842312);
				\draw (7)--(13)(9)--(14)(11)--(13)(6)--(13)(11)--(12)(11)--(14)(7)--(8)(10)--(14)(8)--(9)(10)--(12)(6)--(12);
				\draw (1)--(6) (2)--(7)(3)--(8)(4)--(9)(5)--(10);
				\node[] at  (4.2,2)  {\small $3$};
				\node[] at  (3.6,3.2) {\small $4$};
				\node[] at  (0.7,3.2) {\small $5$};
				\node[] at  (3.6,0.4) {\small $2$};
				\node[] at  (0.7,0.2) {\small $1$};
			\end{tikzpicture}\label{fig:BHB}}
		\subfigure[Pentagon--Box--Pentagon (PBP)] {
			\begin{tikzpicture}[thick, scale=0.60]
				\coordinate (1) at (0.8166,0.000709659);
				\coordinate (2) at (0,1.78333);
				\coordinate (3) at (2.51243,3.9827);
				\coordinate (4) at (5.02661,1.78427);
				\coordinate (5) at (4.20993,0);
				\coordinate (6) at (1.37367,0.733304);
				\coordinate (7) at (0.914117,1.6497);
				\coordinate (8) at (2.51307,3.0549);
				\coordinate (9) at (4.11155,1.65009);
				\coordinate (10) at (3.65257,0.734098);
				\coordinate (11) at (2.51315,1.83164);
				\coordinate (12) at (1.78218,2.32175);
				\coordinate (13) at (3.24285,2.32174);
				\coordinate (14) at (2.51275,0.889231);
				\draw (8)--(13)(6)--(14)(9)--(13)(11)--(13)(7)--(12)(9)--(10)(11)--(12)(8)--(12)(11)--(14)(10)--(14)(6)--(7);
				\draw (1)--(6) (2)--(7)(3)--(8)(4)--(9)(5)--(10);
				\node[above] at  (4.7,1.7) {\small $2$};
				\node[above] at  (2.8,3.5) {\small $3$};
				\node[above] at  (0.3,1.7) {\small $4$};
				\node[below] at  (4.4,0.8) {\small $1$};
				\node[below] at  (0.7,0.8) {\small $5$};
			\end{tikzpicture}\label{fig:PBP}}
		\caption{Genuine three-loop five-point integral families.}
		\label{fig:3l5p}
	\end{figure}

The revolution of multi-loop quantum field theory made it possible for us to move to the next frontier, {\it three-loop five-point} integral families. The results of these three-loop integrals will serve as a foundation for the studies of N$^3$LO amplitudes for two-to-three processes, as well as the theoretical studies of gauge theories. Very recently, one of the integral families was computed in Ref.~\cite{Liu:2024ont}. However, the remaining families, which are significantly more complicated, remain to be computed.

In this article, we present for the first time the complete  results of all massless planar three-loop five-point Feynman integrals, arranged in the so-called PBB, BPB, BHB and PBP families listed in figure~\ref{fig:3l5p}. 
For each family, we construct a complete pure integrals basis which made the differential equation simple~\cite{Henn:2013pwa,Henn:2014qga}. Thanks to the conjectured three-loop planar pentagon alphabet~\cite{Chicherin:2024hes}, and the development of the state-of-the-art integration-by-parts (IBP) program {\sc NeatIBP 1.1}~\cite{Wu:2025aeg}, we analytically reconstruct the differential equation in its canonical form. Furthermore, we determine all boundary values analytically to the order weight six. Thus all these integrals can be efficiently calculated by solving the differential equation, and their symbols are determined as well. 


    This article is organized as follows: In section~\ref{sec:FI}, the massless five-particle kinematics and four planar three-loop five-point families are introduced. In section~\ref{sec:IBP}, the techniques for efficiently performing IBP reduction is discussed. The complete pure basis for each families are given in ~\ref{sec:UT}, which leads to the CDE in section~\ref{sec:CDE}. The boundary value is extracted from CDE in section~\ref{sec:BC}. The complete symbol space will be identified in section~\ref{sec:SB}. In section~\ref{sec:num}, we validate our computation by numeric approach. Finally, section~\ref{sec:sum} provides a summary remarks and an outlook for the future. 



\section{Integral Families}\label{sec:FI}
    The complete set of planar three-loop five particle integral families is displayed in the figure~\ref{fig:3l5p}. In the previous literature~\cite{Liu:2024ont}, the CDE and functions in Euclidean region of the Pentagon-Box-Box (PBB) family shown in figure~\ref{fig:PBB} was determined. 
    
	We consider the five particle scattering in general $D=4-2\varepsilon$ dimensional Minkowski space. The five external momenta $k_i, i\in\{1,2,3,4,5\}$ are on--shell and satisfy the momentum conservation, i.e. $k_i^2=0, \sum_{i=1}^{5}k_i=0$. We adapted the standard parity-even cyclic variables:
	\begin{equation}
		X:=\{s_{12},s_{23},s_{34},s_{45},s_{15}\},~\ s_{ij}=2k_i\cdot k_j \,,
	\end{equation}
	and one parity-odd invariant:
	\begin{equation}
		\epsilon_5=4\mathrm{i}\epsilon_{\mu\nu\rho\sigma}k_1^\mu k_2^\nu k_3^\rho k_4^\sigma.
	\end{equation}	
	For planar scattering at three-loop level, there are only four integral families, which are listed in Figure~\ref{fig:3l5p}.  It is useful to introduce dual coordinates $x_i$ with $k_i=x_i-x_{i-1}$, where the indices take cyclic values $x_6\equiv x_1$, and $ x_{ij} $ denotes $ (x_i-x_j)^2 $. The dual coordinates for three loop momenta $\ell_1,\ell_2,\ell_3$ were denoted by $y_1,y_2,y_3$ with $y_i=x_{i+6}-x_5$. Thereby, the inverse propagators were defined as follows
\begin{align}
	&\left\{\mathscr{D}_i\right\}
	=\left\{
	x_{17}^2,x_{27}^2,x_{37}^2,x_{47}^2,x_{57}^2,\right.
	x_{18}^2,x_{28}^2,x_{38}^2,x_{48}^2,\notag\\
	&\left. x_{58}^2,
	x_{19}^2,x_{29}^2,x_{39}^2,x_{49}^2,x_{59}^2,
	x_{78}^2,x_{79}^2,x_{89}^2
	\right\}\,.\label{tab:18prop}
\end{align}
With these propagators, all integrals families under consideration can be written as:
	\begin{equation}
		I_{\nu_1,\dots, \nu_{18}}(\mathcal{N}) =\ee ^{3\varepsilon \gamma_E}\int 
		\prod_{j=1}^3\frac{ \mathrm{d}^{4-2\varepsilon}\ell_j}{(\mathrm{i} \pi)^{2-\varepsilon}} \frac{\mathcal{N}(\ell,k)}{\prod_{i=1}^{18}\mathscr{D}_i^{\nu_i}},
		\label{eq:intdef}
	\end{equation}
     where $\gamma_E$ is the Euler--Mascheroni constant, $\nu_k$ are the propagator powers and $\mathcal{N}$ is a possible numerator. One can specify a particular family by restricting some of the indices $\nu_i$ to be non--positive,  e.g. $\nu_i\le0$, $i\in\{ 4, 6, 7, 9, 11, 12, 17 \}$ holds for the PBB family in figure~\ref{fig:PBB}.

\section{IBP Reduction via Computational Algebraic Geometry}\label{sec:IBP}
    IBP reductions constitute a major bottleneck in most state-of-the-art multi-loop Feynman integral computations. Consequently, typical three-loop five-point integral family problems remain challenging using IBP reduction tools based on traditional methods ~\cite{Smirnov:2019qkx,Klappert:2020nbg,Lange:2025fba,Smirnov:2025prc}. 
    We found that the public program {\sc NeatIBP}~\cite{Wu:2023upw,Wu:2024paw}. It employs computational algebraic geometry techniques~\cite{Gluza:2010ws,Georgoudis:2016wff,Badger:2013sta,Zhang:2012ce,Boehm:2020zig,Bohm:2018bdy,Larsen:2015ped} to generate IBP equations and helps to address the challenges dramatically. 
    Using this package, we find that the number of master integrals (MIs) for the four families are $316,367,431$ and $734$ respectively. The number of genuine five-point integrals and inequivalent five-point topologies are also identified as listed in table \ref{tab:mis}. Hence, the tennis-court topologies (BHB and PBP) turn out to be considerably more formidable than the ladder-type topologies (PBB and BPB). 
    Thanks to the recent development in the function \texttt{SpanningCut}~\cite{Larsen:2015ped,Bohm:2018bdy} implemented in~{\sc NeatIBP 1.1}~\cite{Wu:2025aeg}, we are able to make the reduction of the tennis-court families feasible. Additionally, each reduction job can be done with remarkable efficiency. The cost of CPU time for each family were also listed in table~\ref{tab:mis}.

	   \begin{table}[htbp]\begin{spacing}{1.1}
			\begin{tabular}{ccccc}\hline
				&PBB&BPB&BHB&PBP\\\hline
				 \# master integrals&$316$&$367$&$431$&$734$\\\hline
	 \makecell[c]{\# genuine \\five-point integrals}&$120$&$164$&$169$&$342$\\\hline
	\makecell[c]{\# independent \\five-point topologies}&$17$ &$19$ &$23$ &$22$\\\hline
    \makecell[c]{\# $10^3$ CPU hour \\ for IBP}
    & $\sim 0.5$ & $\sim 1$ & $\sim 3$ & $\sim 7$\\\hline
			\end{tabular}\end{spacing}\caption{Overview of the integral families.}\label{tab:mis}
		\end{table}

\section{Pure integral basis}\label{sec:UT}

Building and solving differential equations (DE) of the integral basis after IBP reduction is one of the main methods to analytically compute such basis. With a good choice of basis, called pure integral basis, the DE takes a canonical form,
\begin{equation}
    \mathrm{d} \mathbf{I}=\varepsilon \mathrm{d} \tilde{A} \mathbf{I},\ \mathrm{d} \tilde{A}=\sum_{W_i \in \mathbb{A}} a_{ i} \mathrm{~d} \log W_i,
\end{equation}
and can be solved by iterative integration \cite{Henn:2013pwa}. The CDE also tell us about the singularities of the integral families, represented by a set of functions, $W_i$'s which are \textit{symbol alphabets} \cite{Goncharov:2010jf}. 

In this work, we find the pure integral basis for three-loop five-point integrals, so that they can be efficiently calculated by the canonical differential equation method \cite{Henn:2013pwa}.

\begin{figure}[t]
	\centering
	\subfigure[``The shield" with $15$ MIs] {
		\begin{tikzpicture}[thick, scale=0.6]
			\coordinate (1) at (4.22768,2.88363);
			\coordinate (2) at (4.22978,0.850358);
			\coordinate (3) at (0.866046,0);
			\coordinate (4) at (0,1.86391);
			\coordinate (5) at (0.863495,3.73026);
			\coordinate (6) at (3.43296,2.37799);
			\coordinate (7) at (3.43564,1.35433);
			\coordinate (8) at (1.30182,0.845628);
			\coordinate (9) at (1.29909,2.88514);
			\coordinate (10) at (1.85068,1.865);
			\coordinate (11) at (2.32493,1.13269);
			\coordinate (12) at (2.32247,2.59844);
			\coordinate (13) at (0.960194,1.86539);
			\draw (10)--(11)(9)--(13)(7)--(11)(6)--(12)(8)--(13)(6)--(7)(10)--(13)(9)--(12)(8)--(11)(10)--(12);
			\draw (1)--(6) (2)--(7)(3)--(8)(4)--(10) (5)--(9);
            \node[above] at  (4,2.8) {\small $2$};
			\node[above] at  (1.4,3.3) {\small $3$};
			\node[above] at  (0.2,1.8) {\small $4$};
			\node[below] at  (4,0.9) {\small $1$};
			\node[below] at  (1.4,0.5) {\small $5$};
		\end{tikzpicture}\label{fig:shield}}
	\subfigure[``The racehorse'' with $6$ MIs] {
		\begin{tikzpicture}[thick, scale=0.66]
		\coordinate (1) at (1.21609,3.24716);
		\coordinate (2) at (0,1.62486);
		\coordinate (3) at (1.21321,0);
		\coordinate (4) at (4.53534,0.640247);
		\coordinate (5) at (4.53421,2.60616);
		\coordinate (6) at (1.60354,2.40993);
		\coordinate (7) at (0.929232,1.62403);
		\coordinate (8) at (3.77133,1.16754);
		\coordinate (9) at (3.77178,2.07926);
		\coordinate (10) at (2.69543,2.07168);
		\coordinate (11) at (1.60159,0.836438);
		\coordinate (12) at (2.69502,1.17464);
		\draw[] (11)..controls (1.70,1) and (2.2,1.3)..(12);
		\draw[] (11)..controls (1.70,0.8) and (2.3,0.78)..(12);
			\draw (8)--(12)(6)--(10)(7)--(11)(6)--(7)(8)--(9)(9)--(10)(10)--(12);
			\draw (1)--(6) (2)--(7) (3)--(11) (5)--(9) (4)--(8);
            \node[above] at  (4,2.4) {\small $5$};
			\node[above] at  (1.6,3) {\small $1$};
			\node[above] at  (0.2,1.6) {\small $2$};
			\node[below] at  (4,0.9) {\small $4$};
			\node[below] at  (1.6,0.5) {\small $3$};
		\end{tikzpicture}\label{fig:relatively-easy} }
	\caption{New sub-topologies in PBP family.}
	\label{fig:newPBP}
\end{figure}
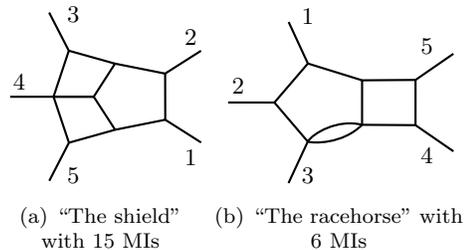

For the integrals with four-point and one off-shell external legs, their pure basis has been deeply investigated in ~\cite{DiVita:2014pza,Henn:2023vbd}. For  genuinely five-point integrals, the number of independent topologies in each family are listed in table~\ref{tab:mis}. There exist partial overlaps between different families--once the three families PBB, BPB and BHB are completed, only three topologies are new in PBP: the top topology PBP with $19$ master integrals shown in figure~\ref{fig:PBP}, and the other two shown in figure~\ref{fig:newPBP}. These topologies contain an exceptionally large number of integrals within a single sector, posing additional challenges for constructing  pure bases.

To construct a pure basis for these genuine five-point integrals, we follow the loop-by-loop leading singularity analysis in momentum or Baikov representation~\cite{Cachazo:2008vp,Henn:2013pwa,Frellesvig:2024ymq}, as well as the chiral numerator from local integrand~\cite{Arkani-Hamed:2010pyv} and $dlog$ integrals based on {\tt Dlogbasis} algorithm \cite{Henn:2020lye}. To systematically construct numerators with unit leading singularities, we employ the algebraic geometric method of module lift in Baikov representation.

Here we list  representative pure integrals on top topology of PBP shown in figure~\ref{fig:PBP}:
			\begin{align}
				\mathcal{N}_4\ &=\frac{\varepsilon(s_{12}-s_{45})}{(1+2\varepsilon)\epsilon_5} \mathrm{G}(\ell_2,\ell_3,k_1,k_2,k_3,k_4)\,,\\
				\mathcal{N}_{13}&=s_{12} s_{34} s_{45}(\ell_2-k_1)^2 (\ell_3-k_{123})^2 \,,\label{eq:pbp4pySecDec}\\
				\mathcal{N}_{17}&=s_{12} s_{45}  (\ell_3+k_5)^2  \big(
								(\ell_2-k_{123})^2  (\ell_3-k_{12})^2\notag\\ 
								&\qquad -(\ell_3-k_{123})^2 (\ell_2-k_{12})^2\big)\,,\\
				\mathcal{N}_{19}&=s_{12}s_{15} s_{45}  
				\big(
				(\ell_2-k_{123})^2 (\ell_3-k_{12})^2 
				\notag\\ &\qquad - 
				(\ell_3-k_{123})^2 (\ell_2-k_{12})^2 
				\big) \,.
			\end{align}
        We abbreviate $k_{1..j} $ as $ \sum_{i=1}^{j}k_i$. The complete pure basis for each family can be found in the auxiliary file \verb|Canonical_Basis.m|, see Appendix~\ref{sec:supp}.

\section{Differential equation and alphabet}\label{sec:CDE}


By taking derivatives of our pure integral basis and performing IBP reduction, we can check semi-analytically that the resulting differential equation is proportional to the regulator $\varepsilon$. Furthermore, with the \textit{three-loop planar pentagon alphabet} conjectured in~\cite{Chicherin:2024hes}, we can reconstruct the analytical CDE for each integral family. As a result, we confirm that the letter alphabet indeed consists of the following fifty-six letters:
\begin{align}\label{eq:letters}
	\mathbb{A}^\text{3-loop}_{\text{P}} =  \mathbb{A}^\text{2-loop}_{\text{P}} \cup& \{\widehat{W}_i\}_{i=1}^{5} \cup \{ \widetilde{W}_{i} \}_{i=16}^{20}\\&
	 \cup \{ \widetilde{W}_{i} \}_{i=41}^{55} \cup \{ \widetilde{W}_{i} \}_{i=76}^{80} \,\notag,
\end{align}where $\mathbb{A}^{\text{2-loop}}_{\text{P}}$ denotes the well-known two-loop planar pentagon alphabet~\cite{Chicherin:2017dob,Gehrmann:2018yef}. Each subset $\{W_i\}_{i=1}^{5}$ forms a cyclic orbit $W_i=\tau^{(i-1)}(W_1)$. Note that the alphabet is closed under the dihedral group, $D_{5}$.

The novel three-loop letters take the following representative forms:
\begin{align}
	\widehat{W}_1&=s_{23} s_{34}-s_{34} s_{45}+s_{45} s_{15}\,,\\
	\widetilde{W}_{16}&=\Delta_4^{(1)}\,,\\
	\widetilde{W}_i&=\frac{q_i-\sqrt{\Delta_4^{(1)}}}{q_i+\sqrt{\Delta_4^{(1)}}}, \ i\in\left\{41,46,51\right\}\,,\\
	\widetilde{W}_{76}&=\frac{q_{76}-\epsilon_5\sqrt{\Delta_4^{(1)}}}{q_{76}+\epsilon_5\sqrt{\Delta_4^{(1)}}}\,,
\end{align}	
where $q_i$ are polynomials in the Mandelstam variables, and $\Delta_4^{(1)}$ is a quartic polynomial given by
\begin{align}
	&\Delta_4^{(1)}=
	s_{15}^2 s_{12}^2+s_{23}^2 s_{12}^2-2 s_{15} s_{23} s_{12}^2-2 s_{23}^2 s_{34} s_{12}
	\notag\\& +2 s_{15} s_{23} s_{34} s_{12} +2 s_{15} s_{34} s_{45} s_{12}
	+2 s_{23} s_{34} s_{45} s_{12}\notag\\&+s_{23}^2 s_{34}^2+s_{34}^2 s_{45}^2-2 s_{23} s_{34}^2 s_{45}\,,
\end{align}
and $\Delta_4^{(i)}=\tau^{(i-1)}(\Delta_4^{(1)})$ for $i=1,...,5$.
The symbol letters can be found using the computer codes \cite{Jiang:2024eaj}. They are given in appendix~\ref{sec:supp}.

Consequently, the CDE for each integral family can be written as:
\begin{align}\label{eq:cderef}
	\mathrm{d}\mathbf{I}_f=\varepsilon \mathrm{d}\tilde{A}_f \mathbf{I}_f, \ \ \mathrm{d}\tilde{A}_f=\sum_{W_i\in\mathbb{A}_f }a_{f;i} \mathrm{dlog}W_i, 
\end{align}
where $f \in \{\text{PBB}, \text{BPB}, \text{BHB}, \text{PBP}\}$ labels the integral family, $a_{f;i}$ are $\mathbb{Q}$-valued constant matrices, and $\mathbb{A}_f$ denotes the subset of letters relevant to each family. 
Explicitly, we find
\begin{align}
	\mathbb{A}_{\text{PBB}}=&\mathbb{A}^\text{2-loop}_{\text{P}},\  \mathbb{A}_{\text{BPB}}=\mathbb{A}^\text{2-loop}_{\text{P}}\cup\{\widehat{W}_i\}_{i=1}^{5} \notag\\
	\mathbb{A}_{\text{BHB}}=&\mathbb{A}_{\text{PBP}}=\mathbb{A}^\text{3-loop}_{\text{P}}\,,
\end{align}
where $\mathbb{A}^\text{3-loop}_{\text{P}}$ is given in eq.~\eqref{eq:letters}. When fixing a specific orientation of the external legs, additional letters may be omitted. Further details are provided in the auxiliary file \verb|Atilde.m|.

By our explicit computation, it is confirmed that the three-loop planar pentagon alphabet is complete, fully capturing the singularities structure of the planar three-loop five-point integrals.
	
\section{Boundary Values}\label{sec:BC}
The boundary values are necessary for solving these CDEs and themselves have interesting properties. The boundary values can be fixed by imposing the absence of spurious singularities~\cite{Henn:2014lfa,Gehrmann:2018yef,Chicherin:2018mue,Henn:2024ngj}. Following the strategy applied to the PBB family (Sec.VII in~\cite{Liu:2024ont}), we determine the boundary constants for the remaining families at the symmetric Euclidean point $x_0=\{-1,-1,-1,-1,-1\}$. Exploiting the physical consistency conditions, the boundary values can be written in a formal form as:
\begin{equation}\label{eq:bcformal1}	\mathrm{I}^{(n)}_i(x_0)=c_i\mathrm{I}^{(n)}_1(x_0)+\sum_{j,k,l}d_{ijk}
	\int_{\gamma_k} \dd \widetilde{A}_{jl}\cdot \mathrm{I}_l^{(n-1)}(x),
\end{equation}
where $\mathrm{I}_1^{(n)}(x)$ denotes a trivial integral computable by Feynman parametrization, and the rational constants $c_i$ and $d_{ijk}$ are fixed by the requirement that all spurious poles cancel. 

Practically, fixing these constants requires integrating the CDE from the base point $x_0$ to the spurious singularities $x_k$ along a path $\gamma_k$. To render the integration expressible in terms of Goncharov polylogarithms~\cite{goncharov2001multiple,goncharov2011multiple,Duhr:2019tlz}, we choose paths along which the alphabet can be rationalized. For the ladder-type families PBB and BPB, which involve a single square root $\epsilon_5$, we use the path $\gamma:s_{i,i+1}=-x/(1-x)^2$~\cite{Gehrmann:2018yef}. 
For the tennis-court families BHB and PBP, the additional square roots $\{\sqrt{\Delta_4^{(i)}} \} _{i=1}^5$ introduce a $\sqrt{x}$ dependence, which can be rationalized by introducing $x\to t^2$:
\begin{equation}\label{eq:bcpath2}
	\gamma:\ s_{i,i+1}=\frac{-t^2}{(1-t^2)^2}.
\end{equation}
Using this parametrization, the three--loop planar pentagon alphabet $\mathbb{A}^\text{3-loop}_{\text{P}}$ reduces to $17$ effective letters.

	Performing the integration explicitly, we obtain all boundary values up to weight six analytically, which involve over $\mathcal{O}(10^6)$ Goncharov polylogarithms. The corresponding numeric values can be obtained to high precision using \texttt{NumPolyLog}~\cite{Gint}. These numerical boundary constants were further validated against independent checks by \texttt{pySecDec}, see section \ref{sec:num}. 
\section{Symbology}\label{sec:SB}
It is important to study the analytic structure of intrinsic multi-loop Feynman integrals through symbology. In this work, symbols of all three-loop five-point planar massless integrals are derived.

The canonical differential equation for each family and each permutation in $D_5$, is obtained from the transformation rules of the alphabet. Then with the weight-zero boundary values, the symbols of the pure basis are calculated up to the weight six.

 We count independent symbols to study the function space. It is difficult to obtain independent symbols directly from all $D_5$ permutations and integral families, since the number of terms in these symbols is huge. We use a divide-and-conquer strategy to determine the symbol space from each family under the all $D_5$ permutation, and then merge to get independent symbols. The number of letters and independent symbols appearing in each weight order are listed in 
Table. \ref{tab:widgets}. The symbol manipulation is powered by finite-field computational tools \cite{Peraro:2019svx,spasm}.
 
\begin{table}[h]
\centering
\begin{tabular}{|l|c|c|c|c|c|c|}
\hline
Weight & 1 & 2 & 3& 4 & 5 & 6\\\hline
Number of letters & 5 & 10 & 20  & 51 & 56 & 56\\
\hline
Number of symbol  & 5 & 20 & 76  & 285 & 1000 & 2220\\
\hline
\end{tabular}
\caption{\label{tab:widgets} Function space of three-loop five-point massless Feynman integrals. For each weight, the number of letters and the number of independent symbols are given.}
\end{table}

Up to weight three, the symbols only consist of two-loop pentagon letters, but starting from weight four, new letters appear. For three-loop five-point integrals, the symbols' first entries only allows $W_1$, ..., $W_5$, while first two entries allow $35$ combinations of letters. Furthermore, $810$ letter pairs never appear in these three-loop Feynman integrals. 

\section{Numerical Evaluation}\label{sec:num}

    Using the CDE and boundary values in section \ref{sec:CDE} and \ref{sec:BC} as input, we can efficiently evaluate each integral family at a Euclidean generic point by numerically solving the CDE with \verb|DiffExp|~\cite{Hidding:2020ytt}. For example, it takes less than $10$ minutes for one core to generate the values for the $734$ pure integrals in PBP at the following generic point,
    \begin{equation}\label{eq:generic_point}
       X:\ \ \left\{-\frac{36}{29},-\frac{32}{23},-\frac{45}{31},-\frac{33}{19},-\frac{25}{18}\right\} \,.
    \end{equation}
     As a comparison, we evaluate one of the top sector integral--the $\mathcal{N}_{13}$ in eq. \eqref{eq:pbp4pySecDec}, by direct Monte-Carlo simulation with the help of the public program \verb|pySecDec|~\cite{Borowka:2017idc,Li:2015foa,Borowka:2018goh}, which takes more than $\mathcal{O}(10^3)$ CPU hours. The results are listed in table \ref{tab:pysol}, which shows the agreement within the margin or error. For the sub-families accessible to the current implementation of \textsc{AMFlow}~\cite{Liu:2022chg,Liu:2017jxz}, we performed additional cross-checks, finding complete agreement. 
    
\section{Summary and outlook}\label{sec:sum}

In this article, we completed the analytic computation of massless planar three--loop five--point Feynman integrals in dimensional regularization, which marks a new milestone in the area of multi-loop multi-leg Feynman integral computation. The complete pure basis for four planar integrals families was explicitly provided. Combined with the recent development of {\sc NeatIBP} program, we achieved the complete canonical differential equation for four families. The analytic boundary values up to the order weight six were also obtained by imposing the absence of spurious singularities.  With these computations, fast numerical evaluation for these integrals can be implemented by solving the differential equation. 
	
Our results provide the fundamental building blocks for  $2\to3$ physical observables at N$^3$LO. They will uplift the future collider physics to a new level of precision. Combined with recent results for two-loop six-point integrals \cite{Henn:2021cyv,Abreu:2024fei,Henn:2024ngj}, the integral families presented here complete the ingredients for N$^3$LO three-particle production processes, which include 3-jet, 3-photon, and mixed light final states, which are relevant for precision phenomenology at the HL-LHC and, in particular, for the high-precision programs of FCC-ee and CEPC. They will also aid in precision tests of the strong coupling constant. 



We  believe that the state-of-the-art techniques presented in this paper, like the IBP reduction and pure integral basis determination via modern computational algebraic geometry, and the alphabet studies, would boost other three-loop and higher-loop precision physics computations in collider physics and gravitational wave physics.

The computations of the Feynman integrals in this article also boost the theoretical studies of gauge theory. In the  recent breakthrough~\cite{Chicherin:2025int}, our results on three-loop five-point integrals enable the bootstrap of Wilson loops with one Lagrangian insertion at the three-loop level in $\mathcal N=4$ super-Yang-Mills theory. 
Furthermore, our results constitute an important step for the direct calculation of three-loop five-point \textit{all-plus} amplitudes in pure Yang-Mills theory, which can be used to test the conjectured dualities between maximally transcendental pieces and Wilson loops with Lagrangian insertion~\cite{Chicherin:2022bov,Chicherin:2022zxo}.

Finally, the planar computation provides an essential stepping stone toward the non-planar three-loop five-point problem.

\section{Acknowledgement}\label{sec:ack}
We are grateful to Johannes Henn for the participance in the early stage of this work and valuable comments on our manuscript.  We acknowledge Longbin Chen, Xuan Chen, John Ellis, David Kosower, Jungwon Lim, Yuanche Liu, Yanqing Ma, Chenyu Wang, Stefan Weinzierl  for enlightening discussions. DC is supported by ANR-24-CE31-7996. YZ is supported by NSFC through Grant No. 12575078 and 12247103.

\section{Supplemental Material}\label{sec:supp}

    To maximize impact and reproducibility, we provide a detailed description of the ancillary files:
    
	\begin{itemize}
		\item \verb|Letters.m|, the three-loop planar pentagon alphabet in eq.~\eqref{eq:letters}.
		\item \verb|<family_name>/Atilde.m|, the CDE matrix $\tilde{A}$ defined in eq.~\eqref{eq:cderef}
		\item \verb|<family_name>/BCnum.m|, numeric boundary values up to finite order with high precision. 
		\item \verb|<family_name>/Canonical_Basis.m|, a pure integrals basis with uniform transcendental weight which leads to CDE.
		\item \verb|<family_name>/SBW.tar.xz|, a compress file for the symbol of the pure integrals. 
        \item \verb|ForbiddenPairs.txt|, the forbidden pairs in the second entry condition, each integer labels the letters in eq.~\eqref{eq:letters}.
    \end{itemize}
	All the ancillary files can be downloaded from the website:	
    \begin{center}
        \url{https://github.com/YongqunXu/All_Planar_3L5P}.
    \end{center}
	
    Moreover, we show the the numerical value of one top sector integral of PBP in eq.~\eqref{eq:pbp4pySecDec}, obtained by \verb|pySecDec| and \texttt{DiffExp} in table~\ref{tab:pysol}.
       \begin{table}[htbp]
        \centering
     \begin{spacing}{1.2}
            \begin{tabular}{ccc}\hline
        $\varepsilon$-order    & \texttt{pySecDec}  & \texttt{DiffExp}\\ \hline
          $\varepsilon^0$      &  $2.000000\pm4.91\times10^{-8}$  & $2$   \\ 
          $\varepsilon^1$      &  $-2.339370\pm6.88\times 10^{-5}$  & $-2.339426$   \\
          $\varepsilon^2$      &  $-16.19419\pm4.36\times 10^{-4}$ & $-16.19414$   \\ 
          $\varepsilon^3$      &  $-17.36078\pm2.22\times 10^{-3}$  & $-17.36262$   \\ 
          $\varepsilon^4$      &  $18.99960\pm1.04\times 10^{-2} $ & $18.98939$   \\ 
          $\varepsilon^5$      &  $79.57537\pm4.03\times 10^{-2} $  & $79.52752$   \\ 
          $\varepsilon^6$      &  $256.3690\pm1.61\times 10^{-1}  $ & $256.2476$   \\ 
          \hline
        \end{tabular}
        \caption{Value of integral \eqref{eq:pbp4pySecDec} evaluated at the Euclidean point \eqref{eq:generic_point} by \texttt{pySecDec} or \texttt{DiffExp}. }
        \label{tab:pysol}
     \end{spacing}
    \end{table}

\bibliographystyle{apsrev4-1}
\bibliography{reference}
\end{document}